\documentclass[12pt]{iopart}

\usepackage{color}
\usepackage{natbib}
\usepackage{iopams}
\usepackage{graphicx}

\begin{document}

\title[Modelling of impaired cerebral blood flow due to gas emboli]{Modelling of impaired cerebral blood flow due to gaseous emboli}

\author{J.P. Hague $^1$, C. Banahan $^2$ and E.M.L. Chung $^{3,4}$}
\address{$^1$ Department of Physical Sciences, The Open University, Walton Hall, Milton Keynes, MK7 6AA, UK}
\address{$^2$ Medical Physics Department, Leicester Royal Infirmary, University Hospitals of Leicester NHS trust, LE1 5WW, UK}
\address{$^3$ Department of Cardiovascular sciences, University of Leicester, Leicester LE1 5WW, UK}
\address{$^4$ Leicester NIHR Biomedical Research Unit in Cardiovascular Disease, Glenfield Hospital, Leicester, LE3 9QP, UK}

\begin{abstract}
Bubbles introduced to the arterial circulation during invasive medical procedures can have devastating consequences for brain function but their effects are currently difficult to quantify. Here we present a Monte-Carlo simulation investigating the impact of gas bubbles on cerebral blood flow. For the first time, this model includes realistic adhesion forces, bubble deformation, fluid dynamical considerations, and bubble dissolution. This allows investigation of the effects of buoyancy, solubility, and blood pressure on embolus clearance. 

Our results illustrate that blockages depend on several factors, including the number and size distribution of incident emboli, dissolution time and blood pressure. We found it essential to model the deformation of  bubbles to avoid overestimation of arterial obstruction. Incorporation of buoyancy effects within our model slightly reduced the overall level of obstruction but did not decrease embolus clearance times. We found that higher blood pressures generate lower levels of obstruction and improve embolus clearance. Finally, we demonstrate the effects of gas solubility and discuss potential clinical applications of the model.
\end{abstract}

\section{Introduction}

Bubbles entering the cerebral circulation can have devastating consequences for brain function, and are most commonly either created {\it de novo} during decompression sickness, or inadvertently introduced to the circulation during cardiovascular interventions, particularly open heart surgery featuring cardio-pulmonary bypass. Previous animal studies have shown that a rapid influx of large volumes of air has potential to be fatal \citep{weenink2012a}, and in humans it is speculated that introduction of smaller bubbles could be a potential cause of post-operative neurocognitive decline following cardiac surgery \citep{ barak2005a}.

Once lodged in the cerebral arteries, bubbles obstruct blood flow causing downstream tissue to be starved of oxygen and pressure changes to be induced in the surrounding vessels. Since brain tissue is particularly sensitive to a shortage of oxygen (hypoxia), arterial blockages can lead rapidly to irreversible biochemical changes and cell death \citep{lipton1999a}. In recent years there has been increasing clinical interest in understanding the relationship between impaired embolus clearance, systemic blood pressure, and cerebral autoregulation \citep{screiber2009a, caplan1998a}. Improved modelling approaches present new possibilities for examining such relationships with a view to guiding strategies to improve patient outcome. 

To attempt to quantify the effects of solid and gaseous emboli of
varying size, we previously developed a minimal model to forecast the
impact of embolisation on blood flow
\citep{chung2007a,hague2009a}. Our previous model featured a very
limited description of embolisation, in which emboli were assumed to
behave as rigid spheres that block vessels of similar size. The model
had a highly simplified description of the fluid dynamics, assuming that
all pressure was dropped over the arterioles, and that flow at each
bifurcation was governed only by the number of arterioles receiving
flow downstream. In this paper, we present a superior model describing
motion of deformable gas bubbles through the vasculature, which
includes the effects of buoyancy, solubility and blood pressure.

There are a number of differences between solid emboli (which may be
almost incompressible and rigid) and gaseous emboli that compress and distort easily
\citep{branger1999a}. These differences have potential to affect the locations in
the vasculature where emboli become lodged and to influence
dissolution time. Gaseous emboli have a propensity to deform
as they move through the vasculature and only block arteries when the
surface area in contact with the walls generates sufficient static
friction (stiction) to oppose motion \citep{suzuki2003a}. Another
significant difference between solid and gaseous emboli is the high
buoyancy of gas bubbles, which has potential to influence embolus
trajectory. This paper goes beyond previous work by introducing a
number of extensions that are needed to properly model the motion of
gaseous emboli through the vascular tree: (1) the emboli are
deformable (2) an approximation for stiction is included, (3) an
iterative fluid dynamical analysis is carried out to estimate the
pressure drop in the whole tree, and (4) an
estimate of buoyancy effects is included.

The aim of this paper is to provide a model of gas embolisation to
help to understand the relationship between fluid dynamical factors, the accumulation of embolic blockages, and impaired bubble clearance for future use in real-time modelling of embolisation. For the first time, we include the effects of blood pressure, embolus buoyancy, bubble deformation, and
a realistic parameterisation for the stiction and dissolution of
emboli. We begin by introducing our model (section \ref{sec:model}). Example results showing the impact of buoyancy, blood pressure and solubility on blockage can be found in section \ref{sec:results}. We then consider possible clinical applications in section \ref{sec:discussion} and provide a summary of our results (section
\ref{sec:summary}).

\section{Model}
\label{sec:model}

\subsection{Bifurcating tree}
\label{sec:tree}

At bifurcations, the radii of parent and daughter vessels are related by the equation,
\begin{equation}
r_p^{\gamma} = r_{dA}^{\gamma}+r_{dB}^{\gamma}
\end{equation}
where we set the bifurcation exponent $\gamma=3$ to be consistent with Murray's law \citep{murray1926a,murray1926b} (for information on recent work, see e.g. \citet{fung1997a,zamir2000a}), $r_p$ is
the radius of the parent vessel, and $r_{dA}$ and $r_{dB}$ are the
radii of the daughter vessels. We assume a symmetric tree with
$r_{dA}=r_{dB}=r_{d}$, taking the radius of the root node to be $0.5$
mm. Therefore, at each level of the tree, $r_d=(2)^{-1/3}r_p$, or in
terms of the level, $i$, of the tree, $r_i=2^{-i/3}r_0$ where $i=0$ is
the root node of radius $r_0$. The tree used in the example simulations presented in this paper comprises 18 levels, extending from a trunk radius of 0.5 mm to 9.84 $\mu$m arterioles. The number of levels and the diameter of the root node can be adjusted to suit a particular clinical application. In the following, the root node of the tree was assumed to be 1 mm in diameter, as this is similar to the minimum diameter vessel that can be imaged using Magnetic Resonance Angiography (i.e. the topology of larger vessels could be modelled based on imaging data). For ultrasound embolus detection applications the radius of the root node should be adjusted to match the radius of the insonated vessel (typically 2.5 mm for Middle Cerebral Artery insonation \citep{chung2006a, banahan2012a}). The model of a bifurcating tree for vessels of less than 1 mm is justified by
analysis of high-resolution images of the human cortex
\citep{cassot2006a}, which reveal that 94\% of branches in the
cerebral vasculature consist of bifurcations \footnote{Of the
  remaining nodes, 4\% are trifurcations, 1\% simple nodes and 0.5\%
  have 4 or more daughters}. Following West, the lengths of the
vessels in our model were taken to be proportional to their radii ($l\propto r$)  \citep{west1997a}.

\begin{figure}
\begin{indented}
\item[]
\includegraphics[width=100mm]{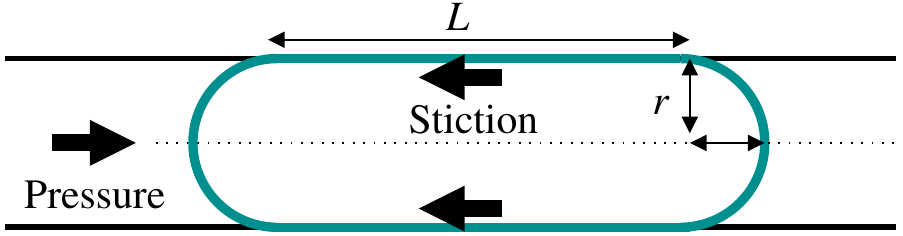}
\end{indented}
\caption{Schematic of a deformable embolus in a vessel, highlighting
  the mechanisms that lead a gas bubble to block an artery. Blood pressure leads to a force, which is opposed by stiction. When the limiting stiction is larger than the force on the bubble from the blood, the artery becomes blocked.}
\label{fig:figure1}
\end{figure}

\subsection{Gaseous emboli}
\label{sec:embolus}

Gaseous emboli differ from solid thrombus and plaque since they are
deformable and rapidly dissolve. To incorporate the effects of deformability and stiction we
assume that a gaseous embolus in a vessel of similar size deforms to a
sausage-like shape (shown schematically in figure
\ref{fig:figure1}). This shape is a common finding in the
cerebral arteries of patients following cardiac surgery, where autopsy
reveals large numbers of sausage-like arteriole dilatations
\citep{moody1990a}. The surface area of the embolus touching the side
of the vessel is computed by correcting for a domed end with the same
radius as the vessel. The length of the cylindrical part of the
embolus is $L=(V_{\rm emb}-4\pi r_{\rm vessel}^3/3)/\pi r_{\rm
  vessel}^2$, and the surface area touching the side is $2\pi r_{\rm
  vessel} L$. 

An embolus will come to rest when the pressure drop over a stationary
embolus is insufficient to overcome stiction, because the force on the embolus generated by the pressure drop, $\pi r_{\rm
  vessel}^2\Delta p$ is less than the limiting force of stiction, $2K\pi r_{\rm vessel} L$, i.e. $\pi r_{\rm
  vessel}^2\Delta p < 2 K (V_{\rm emb}-4\pi r_{\rm vessel}^3/3) /
r_{\rm vessel}$. The coefficient of stiction has been measured to be
$K=10$ Nm$^{-2}$ \citep{suzuki2003a}. The pressure drop for the
stationary embolus is equal to the difference between the pressure at
the bifurcation upstream from the embolus and the capillary network
(since there is no blood-flow in vessels downstream of the embolus the pressure is equal to that of the capillaries).

Based on theoretical
considerations, Branger {\it et al.} (see Eq. 19 in
\cite{branger1999a}) previously developed a model parameterisation to describe
the time that a gaseous embolus of initial volume $V_0$ in mm$^3$ will take to
dissolve,
\begin{equation}
T'_X = 2m_X\pi^{1/3}V_0^{2/3}(2 + X)(4/3 + X)^{-2/3}
\label{eqn:emboluslifetime}
\end{equation}
where $X=L/r$ is the aspect ratio of the embolus given by the length
of the cylindrical section of the bubble, $L$, divided by the radius, $r$, of the artery (see
Fig. \ref{fig:figure1}). The
parameter $m_X$ impacts directly on the lifetime of the bubble for a
specific aspect ratio, $X$, and has been calculated for two aspect ratios
in \cite{branger1999a} to give $m_0 = 97.5$ min/mm$^2$ for a spherical bubble, and $m_{2.6} = 130.9$ min/mm$^2$ where the bubble is cylindrical ($X = 2.6$). We can rearrange this relation to calculate the volume of the bubble at time $t_{emb}$ since it was introduced to the model using:
\begin{equation}
V(t_{emb}) = (4/3+X)\sqrt{\frac{1}{\pi}\left(\frac{T_X-t_{emb}}{m_X(4+2X)}\right)^{3}}
\end{equation}
where $T_X$ and $t_{emb}$ are in mins. Given the limited information regarding $m_X$, we assumed the relation for a spherical bubble ($X=0$) in all subsequent calculations. This represents a best case scenario with fastest dissolve times. These equations describing bubble dissolution have been validated
{\it in-vivo} for small emboli by \cite{branger1999a}, and are thought to be scalable to much larger emboli since the underlying equations represent
gaseous diffusion over a surface and should be approximately valid
for all embolus sizes. When we consider gases other than air, the parameter $m_X$ is scaled accordingly.

\begin{figure}
\begin{indented}
\item[]
\includegraphics[width=100mm]{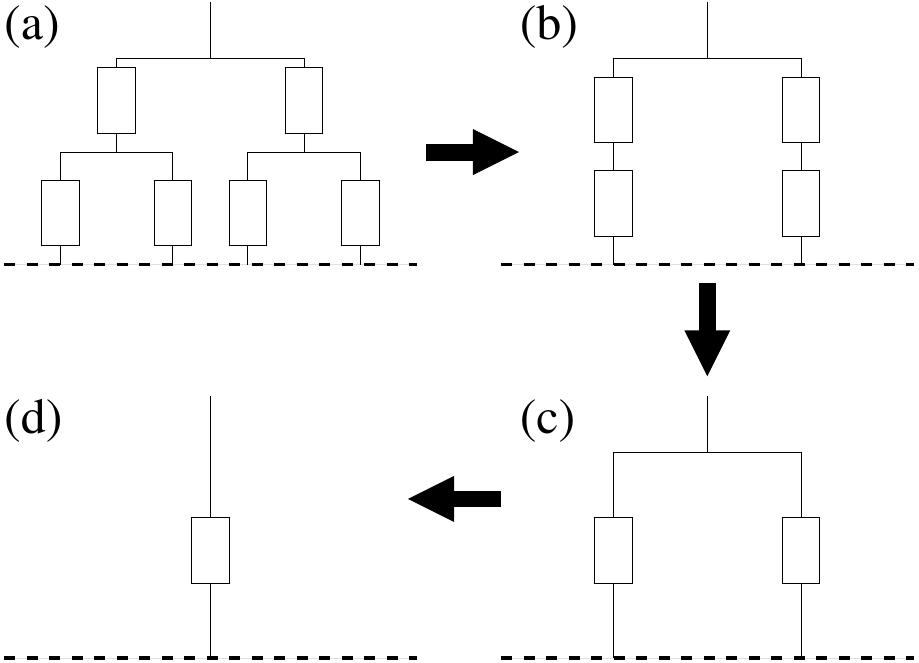}
\end{indented}
\caption{Schematic of the recursive procedure for computing pressures,
  flows and resistances. Pairs of parallel and serial resistances are rewritten as a single resistance recursively through the whole tree until a single resistance remains. The flow through the single resistor can be calculated, and then the procedure is carried out in reverse, calculating pressures, flows and resistances at each level in the tree.}
\label{fig:figure2}
\end{figure}

\subsection{Fluid dynamics and recursive computation}
\label{sec:fluids}

The need to determine the pressures at bifurcations in the
arterial tree is an added complication of dealing with deformable
gaseous emboli, and means that at least a basic fluid dynamical
analysis of flows in the tree is required. The treatment of pressure in
our model also has the important advantage of enabling us to
investigate the theoretical impact of blood pressure changes on embolus
clearance.

When considering the stiction of gas bubbles, it is essential to determine
the pressure difference either side of the embolus. Since pressures need to
be computed whenever an embolus moves, and the bifurcating tree in
the model has a very large number of nodes, it is essential that a
simplified fluid dynamics scheme is used to reduce computation
time. We treat the fluid flow through the tree as
Poiseuille flow, where the pressure drop across a segment is $\Delta p =
Rf$, where $f$ is the flow through the segment and $R$ acts as a
resistance, where $R\propto l/r^4$. Since we assume $l\propto r$,
$R\propto 1/r^3$, and substituting $r_i=2^{-i/3}r_0$, the resistance
at each level is given by $R_i\propto 2^{i}$. The pressure drop and effective
resistance can be treated using an electrical circuit analogue
approach by identifying $\Delta p$ as a potential difference, and $f$
as a current (see e.g. \cite{murray1964a}). Thus, parallel resistances can be rewritten as a single
resistor, and this can be repeated recursively up the tree, working from the
end arterioles to the parent node, to compute all flows and pressures
as an order $N$ operation, where $N$ is the number of vessels in the
tree. This process is summarised in figure
\ref{fig:figure2}, and leads to rapid computations which
are far faster than using matrix inversion or by directly solving
simultaneous equations. 

Panel (a) shows the initial bifurcating tree. In panel (b), the
smallest vessels have been summarised as a single equivalent resistance. In panel
(c) the resistances in series have been simplified, and in panel (d)
the recursive step has been applied to the tree that resulted in
(c). Clearly, this procedure can be applied to any size of bifurcating
tree. Once a single resistance is obtained
for the whole tree, the flow into the tree is calculated, and the
recursion is followed backwards, computing pressures and flows at
every point (e.g. flows and pressures can be computed at step (b) as there is effectively a potential divider in each branch at that
stage. During the recursive procedure, these flows and pressures are
stored, and the stored pressures and flows can be used for all
calculations until any new blockages are introduced or existing blockages are freed as emboli dissolve.

\subsection{Buoyancy}
\label{sec:gravity}

The effects of buoyancy were emulated by introducing a probability weighting, $w_A = (1+A_g\cos(\theta))/2$, related to the orientation of the branches with respect to gravity. $A_g$ is a parameter that varies between 0 and 1, where
$A_g=0$ represents no correction due to buoyancy (where $w=1/2$, as
in the previous version of the model) and $A_g=1$ represents an extreme
correction. This type of weighting is consistent with the
results in \cite{eshpuniyani2005a}, where $\theta$ is the angle between the plane of the bifurcation and the horizontal. We note that this form for the weighting is {\it ad-hoc}, but running the code with $A_g=1$ will demonstrate the essence of the corrections that
are required to describe highly buoyant bubbles. Once the additional weighting factors are introduced, the probability that an embolus travels in direction A at a bifurcation is,
\begin{equation}
P_{A} = w_{A}f_{A} / (w_{A} f_{A} + w_{B} f_{B})
\end{equation}
with $P_B = 1-P_A$, where $f_{A}$ and $f_B$ designate flows in the A and B
directions. $\theta$ is assigned randomly to each bifurcation for each instance of the ensemble at time $t=0$. In future it may be possible to include realistic values for $\theta$ for the cerebral arterial tree based on imaging data or models of angiogenesis.

\subsection{Algorithm}

The algorithm begins by calculating flows, pressures and resistances for an empty tree (using the procedure in section \ref{sec:fluids}). It then proceeds as follows:
\begin{enumerate}
\item On any time step, an embolus may be created in the root node of
  the tree with probability $P_{\tau}\Delta \tau$ with size randomly
  chosen between $0$ and $r_{\rm max}$. Here $P_{\tau}$ is the
  probability per unit time to create an embolus and $\Delta \tau=1s$
  is the length of the time step.
\item All emboli dissolve leading to a reduction in radius during each
  time step according to the parameterisation in
  Sec. \ref{sec:embolus}. Completely dissolved emboli are removed from
  the simulation. If the reduction in radius generates a change in the
  blockage state of the tree, flows and pressures are recalculated.
\item The emboli move according to the following rules:
\begin{enumerate}
\item If the pressure behind the deformed embolus is insufficient to
  overcome stiction it does not move. (See section \ref{sec:fluids})
\item If all arterioles downstream are blocked, the embolus
may not move since there is no flow.
\item If the embolus radius becomes smaller than the current node and
there is flow downstream:
\begin{enumerate}
\item The flows in directions A and B are determined by solving our
  simplified fluid dynamics scheme.
\item The embolus then moves in direction A with probability
$P_A=f_Aw_A/(f_Aw_A+f_Bw_B)$. Otherwise, it moves in direction
B.
\end{enumerate}
\end{enumerate} 
\item If progress of an embolus generates a new blockage, then the pressures and flows are recalculated. At this stage numerical measurements of the state of the tree are repeated.
\end{enumerate}

\subsection{Bubble deformation}
\label{sec:deformation}

\begin{figure}
\begin{indented}
\item[]
\includegraphics[width=120mm]{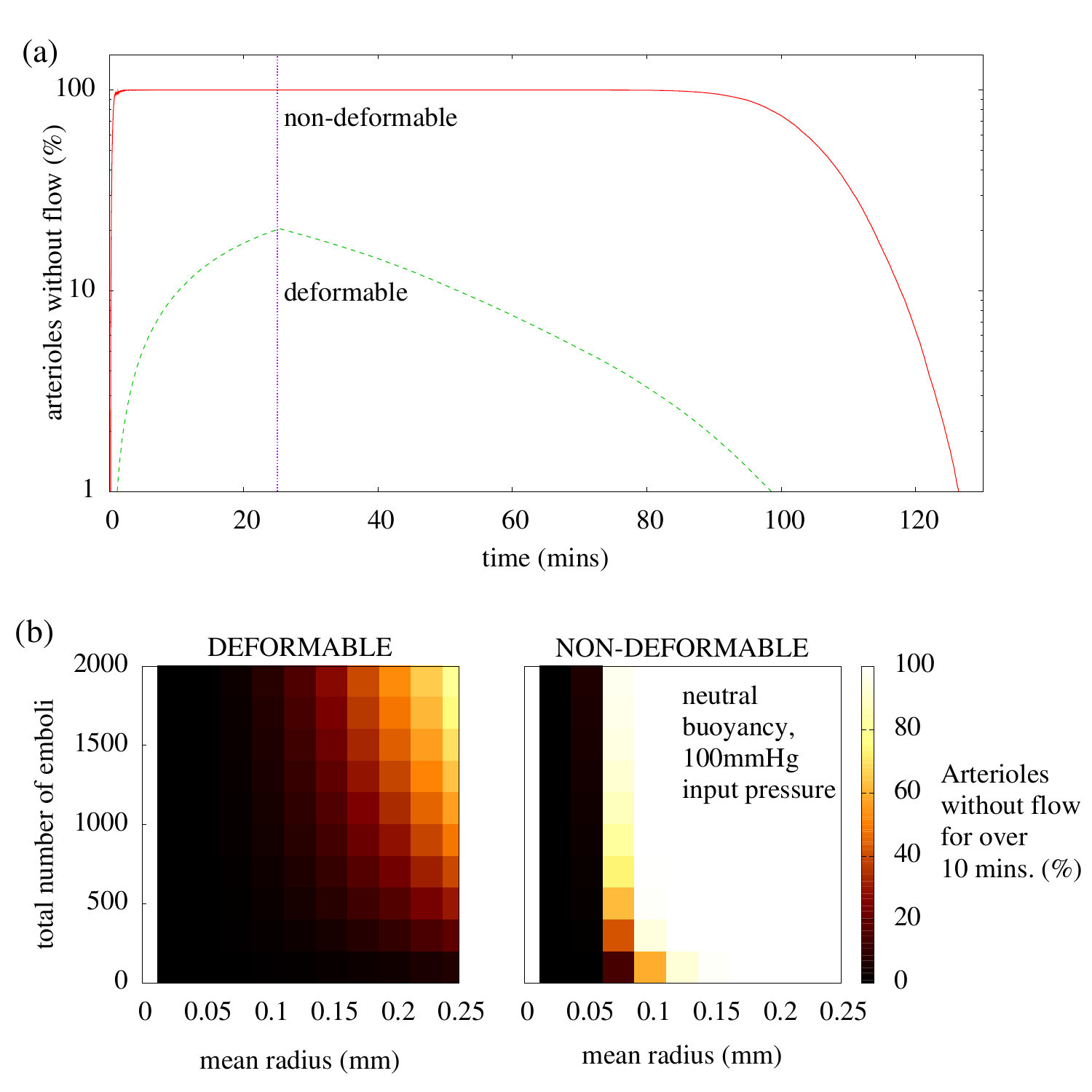}
\end{indented}
\caption{Example simulations comparing the instantaneous percentage of end arteries receiving no flow vs time for hypothetical non-deformable bubbles that remain spherical as they enter the tree compared to more realistic bubbles that deform to the diameter of the vessel. [$P = 100 mm Hg$, $A_g=0$, embolisation rate = 1 emb/s]. (a) When the bubbles are assumed to be spherical and non-deformable the total number of blocked end arterioles rises rapidly and the vascular tree quickly bcomes fully blocked (solid line). Deformation of the bubbles as they enter narrow vessels generates an order of magnitude reduction in the proportion of arterioles without flow (dashed line). The vertical dotted line indicates the time at which embolisation ceases. The total time required for non-deformable bubbles to wash out of the system was longer than for deformable bubbles [$n=1500$, $r_{max} = 0.3$ mm]. Panel (b) shows the proportion of end nodes receiving no blood supply for over 10 mins, plotted as a function of total number of emboli (y-axis) and average embolus radius (x-axis), for deformable and non-deformable emboli. If no blockage is registered, then the cell is black. For non-deformable bubbles larger than approximately 0.1 mm average radius the tree was fully blocked. These results confirm that it is essential to include embolus deformation in the model to avoid overestimation of the proportion of blocked
  nodes.}
\label{fig:figure3}
\end{figure}

To confirm the importance of accounting for bubble deformation within the model  the percentage of blocked end arterioles was estimated for hypothetical non-deformable bubbles that were assumed to remain spherical and become lodged when encountering vessels of equal diameter. When bubbles were not assumed to deform the proportion of blocked nodes rapidly increased to 100\% and the instantaneous percentage of blocked end arteries was significantly overestimated. Example simulations featuring deformable and hypothetical non-deformable bubbles are shown in figure \ref{fig:figure3}.

\section{Results}
\label{sec:results}

Since the model is highly flexible and includes a large number of parameters, the next few sections provide example simulations illustrating the effects of buoyancy, blood pressure, and solubility with all other parameters held fixed. In the absence of clinical data, showers of simulated emboli were generated using a random number generator, which selected embolus radii from a flat distribution ranging from $0$ to $r_{\rm max}$ (the average embolus radius being $r_{\rm max}/2$). For all simulations, embolisation was assumed to commence at $t=10$ s. Emboli were introduced to the tree at randomly generated times centred on a mean embolisation rate of 1 embolus/second until the total number of emboli to be simulated had been delivered. In each simulation the size range of incoming emboli, solubility, buoyancy, blood pressure, and the total number of emboli were varied. To simulate bubbles, all emboli were assumed to be deformable and highly buoyant ($A_g = 1$). The average statistical behaviour of the system was determined from an ensemble of 100 simulations. Outputs of the model include (i) the instantaneous number of end arterioles receiving no flow, (ii) the number of end arterioles without flow for a particular duration (10 mins, 1 hr, or 2 hrs.), and (iii) total time required for washout. Benchmarks of 10 minutes, 1 hr and 2 hrs for obstruction of individual arterioles were chosen to reflect clinically relevant time-scales over which it is thought that neuronal changes might occur. 

\subsection{Buoyancy}

\begin{figure}
\begin{indented}
\item[]
\includegraphics[width=120mm]{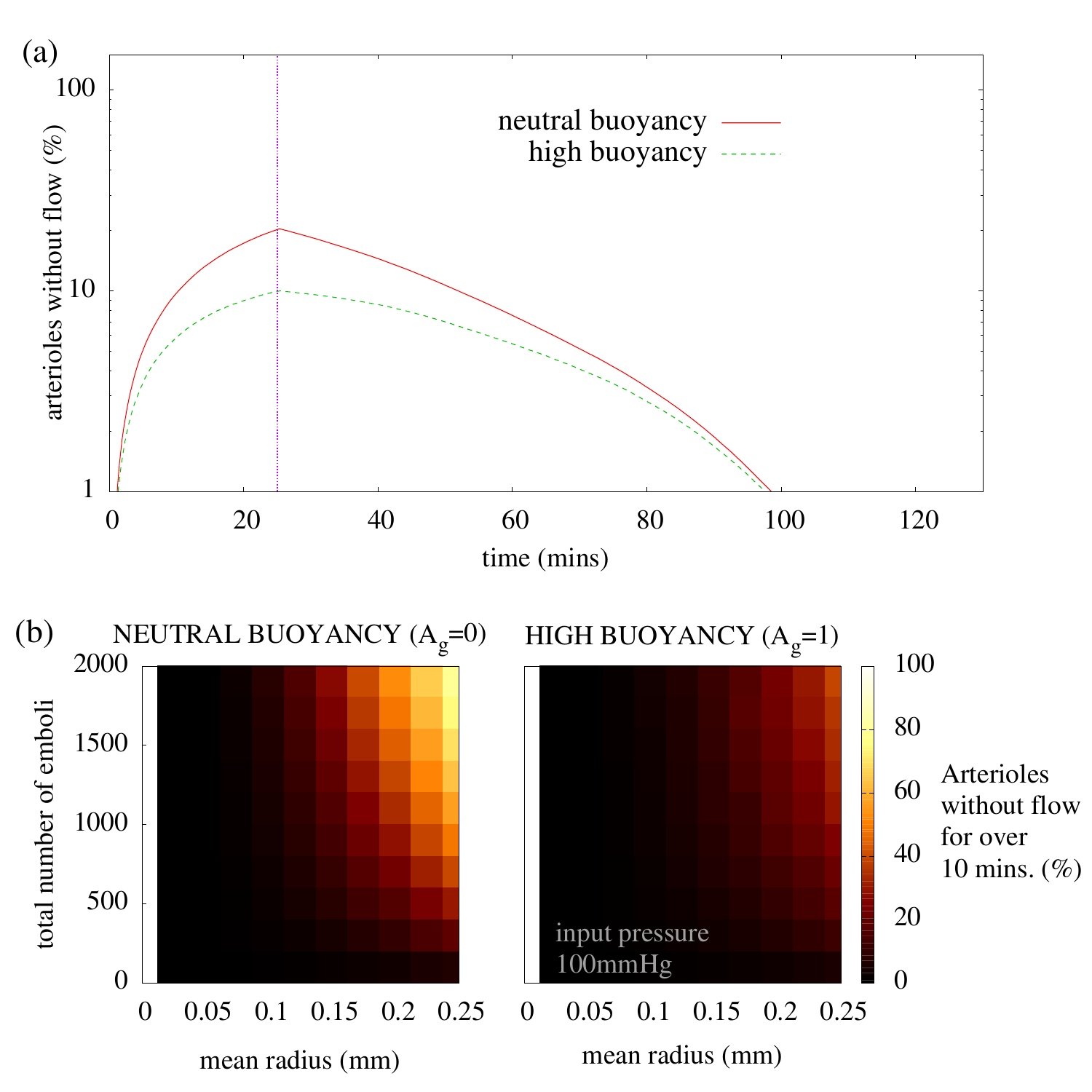}
\end{indented}
\caption{(a) Example simulations comparing the instantaneous percentage of end arteries receiving no flow vs time for neutrally buoyant($A_g = 0$) and highly buoyant ($A_g=1$) bubbles. [$P = 100 $mm Hg, mean embolisation rate = 1 emb/s] (a) After embolisation begins, embolic blockages accumulate and tend toward a dynamic equilibrium level (where similar numbers of emboli are dissolving and leaving the model vasculature as entering). Buoyancy reduces the proportion of end arterioles without flow because certain paths through the tree become more probable when emboli are buoyant. In the washout phase, emboli dissolve without being replaced and the percentage of end arteries experiencing impaired blood flow gradually returns to zero. However, the total time taken for embolus clearance is not reduced by the effects of buoyancy, which indicates regional intensification. [$n=1500$, $r_{\rm max}= 0.3$ mm] (b) shows 2D plots illustrating the effects of buoyancy on the number of individual end arterioles that received no flow for over 10 mins as a function of total number of emboli and average radius.}
\label{fig:figure4}
\end{figure}

Example simulations presented in figure \ref{fig:figure4} illustrate the effects of buoyancy on the number of end arterioles receiving no flow and time required for embolic washout. In panel (a) emboli are randomly generated with an average radius of $r_{\rm av}=0.15$ mm (randomly selected from a flat distribution from 0 to $r_{\rm max}$=0.3 mm). The period of embolisation features 1500 emboli introduced at an average rate of one embolus every second over a period of approximately 25 mins. For reference, an average 0.15 mm radius bubble in this simulation is estimated to take 27.6 minutes to dissolve while the largest 0.30 mm radius bubble takes 110 minutes. As in our previously published model, after embolisation
begins the proportion of end arterioles without flow tends toward a dynamic
equilibrium level \citep{chung2007a}. However, here we are also interested in embolus clearance, so we allow embolisation to cease and emboli to wash out. Following embolisation, existing emboli dissolve and blockages are
cleared.

The effects of buoyancy can be seen to slightly reduce the proportion of
blocked end arterioles due to some paths through the tree becoming more
probable than others. This decreases the total number of obstructed end arterioles but was not found to reduce the time for embolic washout. This effect is expected to be especially pronounced for highly
buoyant emboli ($A_g=1$) since buoyancy leads to preferred paths
through the vasculature and a higher probability of emboli obstructing the
same nodes.

\subsection{Blood pressure}
  
\begin{figure}
\begin{indented}
\item[]
\includegraphics[width=120mm]{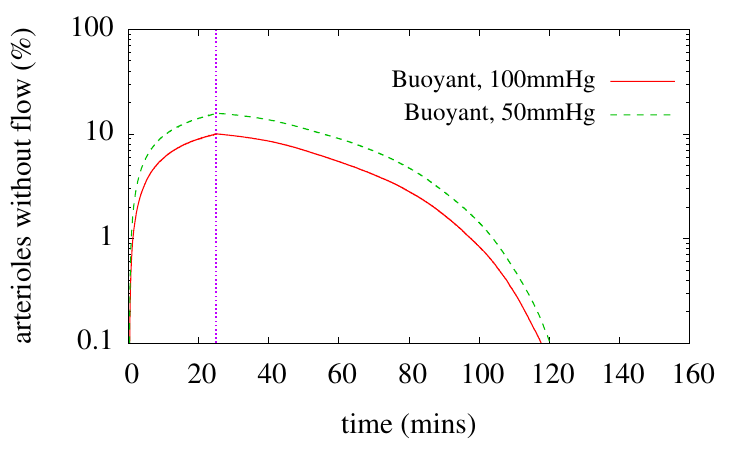}
\end{indented}
\caption{Instantaneous blockages for input pressures of 50 and 100 mm Hg with all other parameters held constant.  [$A_g=1$, $n=1500$, $r_{\rm max} = 0.3$ mm, embolisation rate = 1 emb/s]}
\label{fig:figure5}
\end{figure}

To demonstrate the impact of blood pressure changes on the total number of end arterioles without flow, simulations were performed for  input pressures of 50 and 100 mmHg. The total instantaneous percentage of end arteries receiving no flow decreases with increasing pressure, see figure \ref{fig:figure5}. 

\begin{figure}
\begin{indented}
\item[]
\includegraphics[width=120mm]{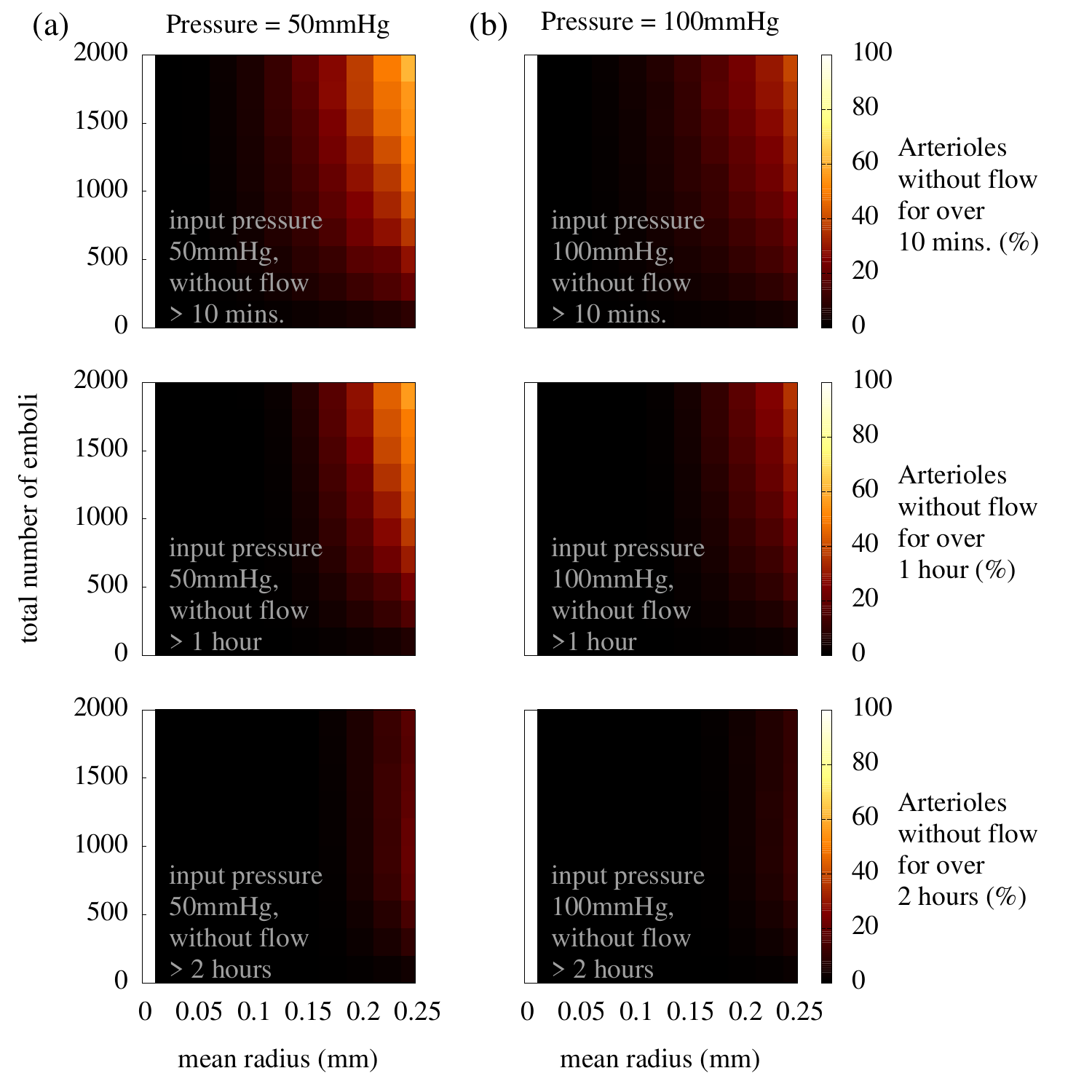}
\end{indented}
\caption{The proportion of end arterioles receiving no supply for 10
  minutes, 1 hour, and 2 hours for buoyant emboli of varying number
  (y-axis) and size (x-axis). The pressure at  the root node is 50 mmHg in the l.h.s. panels and 100 mmHg on the r.h.s. Emboli were introduced at a rate of 1 emb/s and $A_{g}=1$.}
\label{fig:figure6}
\end{figure}

Figure \ref{fig:figure6} shows the effects
of doubling the input pressure  on the percentage of end arterioles that were obstructed for longer than 10 minutes, 1 hour and 2 hours. In all simulations, bubbles were assumed to be deformable and highly buoyant. These results suggest that the duration of embolic blockages due to  gas emboli can be expected to fluctuate with changes in blood pressure. We find that increased blood pressure leads to a reduction in both the total number and duration of embolic blockages.

\subsection{Solubility}

\begin{figure}
\begin{indented}
\item[]

\includegraphics[width=120mm]{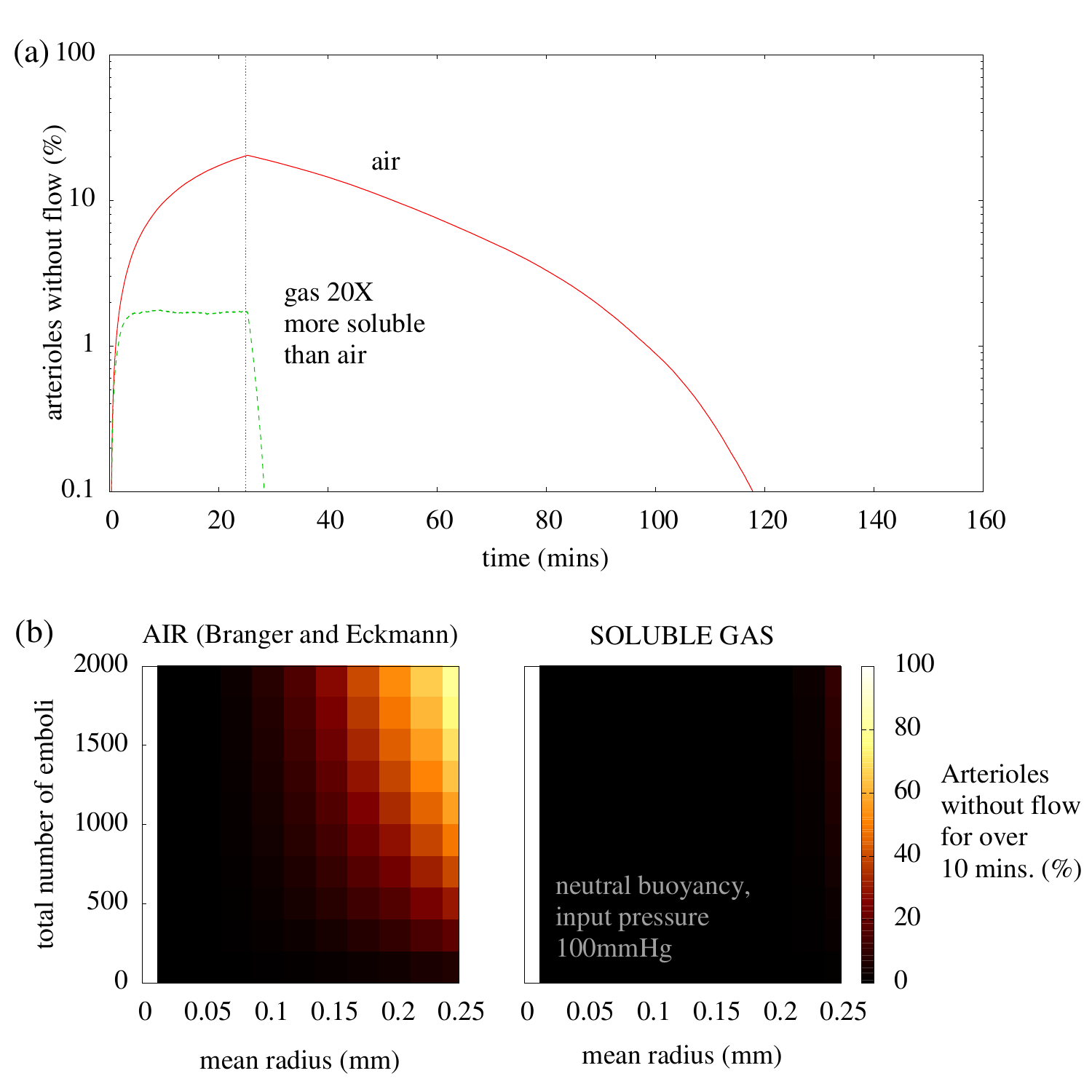}
\end{indented}
\caption{(a) Example simulations showing the accumulation of embolic blockages and washout phase for air compared to a gas that is 20 times more soluble. Increased solubility reduces the instantaneous percentage of arterioles receiving no flow and blockages clear almost instantaneously once embolisation ceases.  [$P = 100$ mmHg, $A_g=1$, $n=1500$, $r_{\rm max} = 0.3$ mm, embolisation rate = 1 emb/s] (b) Percentage of model terminal arterioles without flow for at least 10 minutes for air compared to a gas that is 20 times more soluble. The increase in dissolve rate leads to a dramatic reduction in blockage. Even for simulations featuring bubbles with radii approaching that of the root node (~0.5 mm max radius) virtually no end arterioles remained obstructed for longer than 10 minutes.  [$P = 100$ mmHg, $A_g=1$, embolisation rate = 1 emb/s]}
\label{fig:figure7}
\end{figure}

Finally, we investigated the effects of altering the dissolve rate of emboli, figure \ref{fig:figure7}. Although the simulations performed as part of this study are purely hypothetical, in practice a faster dissolve rate could be achieved by replacing air with a more soluble gas such as CO$_2$. Increasing the dissolve rate of emboli by a factor of 20 was found to significantly reduce the duration of blockages. Figure \ref{fig:figure7}(a) shows the instantaneous percentage of arterioles receiving no flow for 1500 emboli incident on the tree at an average rate of 1 embolus per second, with an average radius of $0.15$ mm (0.3 mm maximum radius). The increase in dissolve rate significantly decreased the total number of blocked end arteries and was observed to have a striking impact on the total time taken for emboli to wash out of the system. Panel (b) shows the reduction in arterioles receiving no flow for over 10 minutes.

\section{Discussion}
\label{sec:discussion}

Despite air embolism representing an important clinical problem, surprisingly
little theoretical modelling has previously been undertaken to
try to quantify the impact of bubbles on
cerebral blood flow. The current study contributes to our understanding of the accumulation of emboli, interplay between blood pressure and bubble properties,
and the process of embolic washout. To the best of our knowledge, our model is the first to incorporate realistic bubble deformation, solubility, buoyancy, and stiction.

Figure \ref{fig:figure3} illustrates the importance of modelling
bubble deformation. In the absence of deformation the total instantaneous
percentage of arterioles without flow rises rapidly and is
significantly overestimated. Assuming hypothetical non-deformable spherical
bubbles the arterial tree quickly reaches saturation. 

In the results section we presented results for deformable emboli and investigated the effects of buoyancy, blood pressure and solubility. The effects of buoyancy of the bubbles can be seen in  figure \ref{fig:figure4} to slightly reduce the total number of end arteries receiving no flow without decreasing the total time required for emboli to completely wash out of the tree. 

An interesting finding of our model was the relationship
between embolisation dynamics, washout of bubbles, and blood
pressure. An increase in blood pressure was found to decrease the proportion of blocked arterioles (figures \ref{fig:figure5} and \ref{fig:figure6}) and appeared to slightly reduced the time required for bubbles to clear. Our results suggest that an increase in blood pressure may assist in
forcing bubbles through the vascular tree more rapidly, thereby
reducing the total number of affected arterioles. This is based on the
principle that higher pressures push gaseous emboli further into the
tree, which reduces the total area of the vasculature that will be affected, (since the number of nodes downstream from a blockage halves at each bifurcation). A similar effect would also be expected to be associated with a decrease in stiction through the introduction of surfactants. This method for reducing the impact of air emboli has previously been tested in animals but not in humans \citep{barak2005a}. We note that a change in blood pressure will slightly modify the solubility here \citep{branger1999a} (for example change in blood pressure from 100mmHg to 50mmHg will decrease the total pressure in the blood - the sum of atmospheric pressure and blood pressure - by approximately 5\%, leading to similar proportional changes in the dissolve rate).

Gas solubility had an impact on both the instantaneous number of arterioles without flow and bubble clearance time; more rapid dissolve times led to faster embolus clearance, figure \ref{fig:figure7}. The dramatic reduction in the total number of end arterioles receiving no flow suggests that if bubbles were formed from a much more soluble gas (such as CO$_2$) both the percentage of arterioles without flow and clearance time would dramatically reduce. At the faster dissolve rate no arterioles experienced blockages lasting longer than 20 mins and the washout period was negligible. This finding is consistent
with {\it in vivo} research conducted on animals showing that the
fatal dose of injected arterial gas is around 50 times higher for
infusion with CO$_2$ than for air \citep{moore1940a}.  Our results are
also consistent with ultrasound embolus detection studies during cardiac surgery which show
that flooding of the operative area with CO$_2$ reduces the number of emboli detected in the
cerebral bloodstream by $~75\%$ \citep{svenarud2004a}. It is hoped that combining our simulations with clinical detection and sizing of bubbles,\citep{banahan2012a}, will enable us to help answer some of these ongoing research questions in future work.

As the model remains highly simplified it retains a number of limitations. Firstly, our tree is completely symmetric and bifurcation angles were randomly assigned at the start of each simulation. We are currently in the process of growing anatomically realistic cerebral vasculatures {\it in-silico} and hope to have an opportunity to combine these with
patient specific clinical imaging data, computational forecasting, and
multi-scale modelling of diffusion and biochemical interactions in
future work. In our model, gaseous emboli do not split at bifurcations or
coalesce. Bubble splitting is likely to affect embolus
trajectory at bifurcations angled within, or close to, the horizontal
plane. In our model of bubble dissolution, we do not include a treatment of the change in dissolution time due to bubble deformation. Also, we assume that the forces required to lodge and dislodge the bubble are equal. So far, we have only considered steady flow conditions (e.g. during cardiopulmonary bypass), however, we expect to be able to include windkessel equations to also describe pulsatile flow in future models. In view of these limitations, the results of the current study are not intended to inform clinical practice, but rather, may highlight areas of interest for further study.

Although the total instantaneous percentage of end
arteries without flow is of limited clinical value, we also investigated the
proportion of arterioles that theoretically received zero flow for
longer than a pre-defined cut-off time (e.g. 10 mins, 1 hr, 2 hrs). 
Further work is required to relate
the duration of impaired perfusion to models of cell death describing
the timescale of neuronal changes, reversible ischaemia, and
irreversible tissue damage \citep{lipton1999a}. This is likely to
require the development of multi-scale models of cell death, which combine
cellular biochemical interactions, solution of localised diffusion
equations, and realistic embedding of the vasculature within brain
tissue. Further work will also be required to incorporate the impact of
haemodilution (which is highly relevant in a cardiac surgery setting), collateral flow, and 
cerebral autoregulation.

By combining a realistic description of the deformation and
dissolution of bubbles with a symmetric model of the cerebral arterial
tree and a simplified description of fluid dynamics, we believe that
we have succeeded in qualitatively understanding the likely extent and
duration of cerebral embolic blockages due to gas bubbles. Given
recent advances in bubble sizing \citep{banahan2012a}, and knowledge
of patient-specific anatomy and physiology, we anticipate that it will
soon be possible to make intra-operative predictions of the impact of
embolisation during surgery.

\section{Summary}
\label{sec:summary}

This paper describes a Monte-Carlo simulation used to model the motion
of gaseous emboli through the cerebral vasculature. Our model improves
on previous research by modelling deformable gas bubbles and 
includes realistic stiction effects, fluid dynamical considerations (including
blood pressure), and a basic description of embolus buoyancy. We show
that our model can be used to investigate the dynamic nature of
cerebral embolisation and to estimate the duration of impaired blood flow
in individual arterioles over time. Using this model it becomes
possible to examine the effects of input
pressure, embolus composition, buoyancy, stiction, size, and
embolisation rate on cerebral blood flow. We found that deformation of
gas bubbles is crucial for quantifying embolic obstruction in response
to blood pressure.  We also give examples of the potential of our
model for investigating factors that influence the impact of gaseous emboli during surgery. Buoyancy effects tend to influence embolus trajectory and generate regional intensification of
blockages.  Since the accumulation of embolic
blockages is partly dependent on blood pressure, maintenance of higher blood pressures might improve embolic
washout by rapidly forcing bubbles through the vasculature, and may be worth further study.  We also confirm that replacement of air with a more soluble gas could theoretically eliminate the risk of gas embolism.

\ack

EMLC is a British Heart Foundation Intermediate Basic Science Research Fellow (FS/10/46/28350). CB and EMLC acknowedge support for this study from the Leicestershire, Northamptonshire and Rutland Comprehensive Local Research Network (CLRN). This study is part of the research portfolio supported by the Leicester NIHR Biomedical Research Unit in Cardiovascular Disease. We acknowledge useful discussions with Jonathan Keelan and David Marshall.



\bibliography{gasembolus}

\end{document}